\documentstyle[prb,epsfig,aps]{revtex} 
\begin{document}
\draft

\title{Determination of the (3$\times$3)-Sn/Ge(111) 
structure by photoelectron diffraction.}

\author{
L. Petaccia,$^{a}$  
L. Floreano,$^{a,}$\footnote{Corresponding Author:
Luca Floreano,
Surface Division,
Laboratorio TASC-INFM,
Basovizza, SS14 Km 163.5
I-34012 Trieste,
Italy.
E-mail: floreano@sci.area.trieste.it}
M. Benes,$^{a,b}$
D. Cvetko,$^{a,c,d}$ 
A. Goldoni,$^{c}$
L. Grill,$^{a}$
A. Morgante,$^{a,b}$ 
A. Verdini,$^{a}$ and 
S. Modesti$^{a,b}$
}

\address{
$^{a}$ Laboratorio TASC, Istituto Nazionale per la Fisica della Materia, 
Basovizza SS14 Km 163.5, I-34012 Trieste, Italy}

\address{
$^{b}$ Department of Physics, University of Trieste, Via 
Valerio 2, I-34100 Trieste, Italy}

\address{
$^{c}$ Sincrotrone Trieste SCpA, Basovizza SS14 Km 163.5, I-34012 Trieste, 
Italy}

\address{
$^{d}$ Jo{\v z}ef Stefan Institute, Department of Physics, 
Ljubljana University, Ljubljana, Slovenia}

\narrowtext
\onecolumn

\maketitle
\begin{abstract}

The bonding geometry of Sn on Ge(111) has been quantitatively 
determined for the $(3 \times 3)$ phase at 150~K. Energy scan 
photoelectron diffraction of the Sn {\it 4d} core levels
 has been used to independently measure the 
bond length between Sn and its nearest neighbour Ge atoms and the 
vertical distance between Sn and the Ge atom beneath. This latter 
distance is found to be $\sim$~0.3~\AA~ larger for one Sn atom out of 
the three contained in the lattice unit cell. The bond lengths and 
the bond directions, obtained by the angular scans, are found to be 
practically the same for the three Sn 
atoms within $\pm$~0.03~\AA~ and $\pm$~3$^{\circ}$, respectively. 
The three nearest neighbour Ge atoms thus partially follow the Sn 
atom in its vertical ripple.
\end{abstract}
\vspace{1cm}

\pacs{PACS numbers: 61.14Qp, 79.60Dp, 68.35Bs}

\narrowtext
\twocolumn

\section{Introduction}

At a coverage of about 1/3 monolayer (ML), Sn and Pb deposited on Ge(111) 
below $\sim$~550~K form a metastable 
$(\sqrt 3 \times \sqrt 3)$R30$^{\circ}$ phase, 
where the Sn or Pb atoms occupy T$_{4}$~ sites above the Ge 
lattice. This phase continuously and reversibly transforms 
into a $(3 \times 3)$ 
one, upon cooling below $\sim$~200 K. This new phase was first 
observed by scanning tunneling microscopy (STM) on the Pb/Ge and 
Sn/Ge systems
and attributed to the 
manifestation of a surface charge density wave 
(CDW).\cite{carpinelli,carpinelli2,scandolo} The 
nature of this phase transition is 
still an open issue, also because of the controversal  determination 
of the atomic 
structure of the 
two phases.\cite{baddorf,bunk,mascaraque} The low temperature 
$(3 \times 3)$  
phase is formed 
by three inequivalent Sn atoms per unit cell. X-ray diffraction (XRD) 
and low energy electron diffraction (LEED) 
experiments now consistently point to an atomic structure where one 
Sn atom out of three protrudes above the surface.\cite{bunk,zhang} 
A vertical ripple 
of $\sim$~0.3~\AA~ is thus observed between the Sn atoms of the surface 
unit cell. Within the CDW model the inequivalence 
between the Sn atoms should disappear in the room temperature (RT) 
$(\sqrt 3 \times \sqrt 3)$ phase. This is consistent with the most recent
experiments that indicate equivalent Sn atoms at the same height 
level in the adatom layer at RT.\cite{bunk,zhang}

X-ray photoemission spectroscopy (XPS) experiments on the 
$(3 \times 3)$ 
phase indeed show two Sn {\it 4d} core level doublets split by 
$\sim$~0.4~eV, with an intensity 
 ratio of approximately 2:1 between the two components 
 (A and B, majority 
and minority components, respectively),\cite{lelay,uhrberg,avila99} thus 
indicating two different kinds of Sn atoms. This spectrum remains 
practically 
 unchanged  throughout the 
phase transition.\cite{lelay,uhrberg,avila} 
This opens the way to alternative explanations, such as dynamical 
fluctuations\cite{avila} and
 order-disorder phase transitions,\cite{uhrberg} where the 
local structure of the two phases is the same. 
Very recently the 
strong influence on the phase transition
 of small defect concentration has been put in 
 evidence.\cite{melechko,weitering} 
Recently published 
XPS studies\cite{kidd} suggest that, at exactly 1/3~ML coverage, 
the Sn {\it 4d} spectrum is different from that 
reported in the literature up to now. In particular the spectrum 
 would show an intensity ratio between the A and B component 
 appreciably different from that reported in previous works and would 
 have an additional component in the high temperature
  $(\sqrt 3 \times \sqrt 3)$ phase. These findings were used to 
  support the CDW model.

It is therefore important to determine the exact lineshape of the Sn $4d$ 
spectrum and to measure the  $(3 \times 
3)$ structure by XPS experiments on the single Sn core level 
components in order to check its consistency with that 
obtained by XRD and to identify the bonding sites of the atoms 
contributing to the majority and minority components. 
We have thus performed photoelectron diffraction 
(PED) measurements of the Sn {\it 4d} core levels at the ALOISA 
beamline (Elettra Synchrotron). 
The wide variety of scattering geometries accessible at the 
ALOISA end station allowed us to measure separately  the bond 
direction and length between the Sn adatom and its three nearest neighbour 
Ge atoms (Ge$_{nn}$) and the Sn vertical distance to the 
next nearest neighbour
Ge atom beneath (Ge$_{nnn}$). 
In a previous paper we qualitatively reported on the preliminary 
experimental results, i.e. the substantial equivalence of the 
Sn--Ge$_{nn}$ bond angle and length for both  A and B type Sn atoms, which 
mainly differ for their vertical height Sn--Ge$_{nnn}$.\cite{floreano}
Here we present the quantitative determination of the bonding 
geometry of Sn atoms in the $(3 \times 
3)$ unit cell, as obtained by fitting to multiple scattering calculations 
(MSCD)\cite{mscd} the PED data obtained in a more extended energy range 
and with a better statistics. 
 We attribute each of the two 
components of the Sn {\it 4d} photoemission spectrum to a specific 
bonding geometry of the two inequivalent types of Sn atoms. 
To our knowledge, such a direct correlation was never extracted before.
The rippled structure obtained in the present work is in good 
agreement both with previous XRD determinations\cite{bunk,zhang} and 
with recent density functional calculations (DFT) 
performed in the local density approximation with gradient correction
(LDA-GC)\cite{degironcoli} and in combination with local-orbital 
(LDA-LO) methods\cite{ortega}.

\section{Experimental set-up and sample preparation}

\begin{figure}
	\centering
	\includegraphics[width=.44\textwidth]{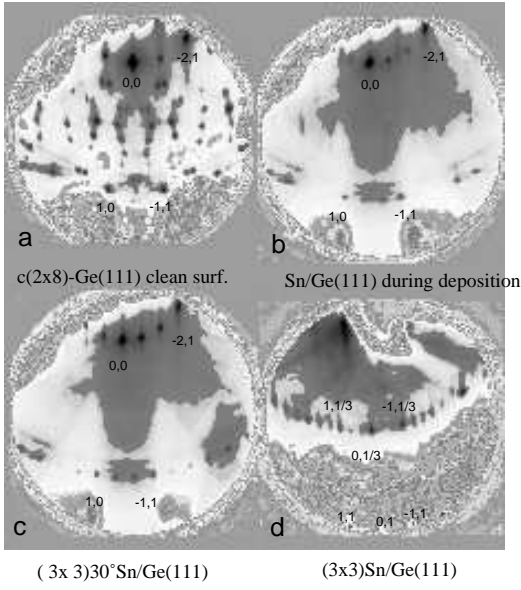}
\caption{ RHEED patterns (15 keV) taken from the clean surface at Room 
Temperature (panel 
$a$) along the $<112>$ direction, spots arise from the c$(2 \times 8)$ 
structure; in panel $b$ the pattern is taken at T$_{s}=500$~K after 
about 4/5 of the 
total evaporation time, the $(\sqrt 3 \times \sqrt 3)$ spots are 
clearly visible, while the half-integer ones are very weak and 
streaky; the evaporation is stopped in correspondence of the panel 
$c$ (T$_{s}=500$~K), 
i.e. when the half-integer peaks completly disappear; the RHEED 
pattern taken along the $<110>$ direction displays the $(3 \times 3)$ 
symmetry upon cooling down to 150~K (panel $d$).
}
\label{figRHEED}
\end{figure}

The present experiment has been performed in the UHV end station of 
the ALOISA beamline\cite{alo99,aloweb} at the Elettra synchrotron 
(Trieste). Details about the experimental setup and sample cleaning 
are given elsewhere.\cite{floreano} We have used a Reflection High 
Energy Electron Diffraction (RHEED) system to monitor in real time 
the preparation of the $\alpha$-phase of Sn on Ge(111).
We have 
followed a procedure slightly different from the standard one reported 
in the literature. Sn is usually evaporated on samples held at RT
 and successively annealed to T$_{s} \sim$~500~K. In this conditions, 
only a diffuse diffraction pattern is observed during deposition and 
the surface quality can be only judged a posteriori. Therefore the same 
surface coverage may not be easily reproduced, being subject to any 
unstability of the
evaporation cell.  By He atom scattering (HAS) experiments with high 
angular resolution, we have 
seen that the best surface quality (i.e. narrowest diffraction peaks 
of both the $(3 \times 3)$ and $(\sqrt 3 \times \sqrt 3)$ phases) 
is obtained when depositing on the surface kept close to 500~K.\cite{has} 
 At the ALOISA end-station we have used 
the RHEED system (which is a long range order probe similar to HAS) 
to monitor the diffraction pattern along the $<112>$ surface direction 
while depositing at T$_{s} \sim$~500~K. The $c(2\times 8)$ spots soon 
disappear leaving only a streaky $(2\times 2)$ pattern.\cite{dicenzo}
 After 2/3 
of the total exposure used in our work, the spots of the 
$(\sqrt 3 \times \sqrt 3)$ 
start to appear (see panel $b$ of Fig.~\ref{figRHEED}), but the 
deposition is just stopped when the half-integer streaky peaks 
disappear (see panel $c$ of Fig.~\ref{figRHEED}). The surface is shortly 
flashed at about 530~K and then cooled down.
Higher flashing temperature irreversibly 
leads to a stable $(7 \times 7)$ phase, as previously 
reported \cite{ichikawa}. Shorter exposures never lead to the 
complete disappearance of 
the  $(2\times 2)$ pattern.
After this procedure sharp spots of the  $(\sqrt 3 \times \sqrt 
3)$R30$^{\circ}$ phase are observed along the $<112>$ direction. 
To detect the spots characteristic of 
the $(3 \times 3)$ phase, the RHEED pattern was observed along the 
$<110>$ direction, as shown in the $d$ panel of Fig.~\ref{figRHEED}, 
which 
is taken at T$_{s} \sim$~150~K. 
 We have seen 
that no degradation of the photoemission spectra nor of the RHEED 
patterns occurs for at least 10 hours, when the sample is left inside the 
experimental chamber. 
In any case, a new surface was always prepared after 10 hours. 

The whole set of spectroscopical 
data hereafter shown has been taken with constant electron energy 
resolution of 170~meV. The photon energy resolution has been always 
$\leq 150$~meV. The overall instrumental resolution is thus 
$\leq 225$~meV.

\section{Results and Discussion}

\begin{figure}
	\centering
	\includegraphics[width=.46\textwidth]{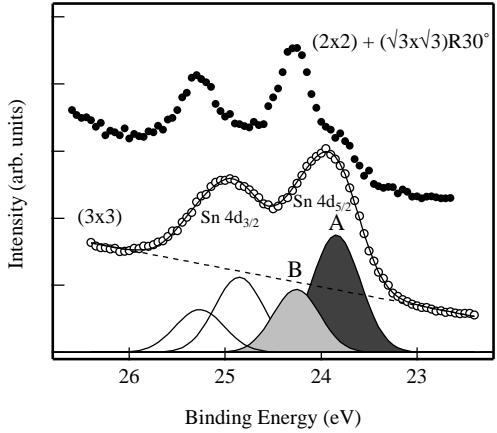}
\caption{ The Sn 4d photoemission spectrum taken at a photon energy 
of $\sim$~210~eV and an overall resolution of 220~meV, normal emission, 
TM polarization. 
The experimental data (open circles $\circ$) are shown with the fit (full 
line) to four gaussians
and a quadratic background (dotted line). 
The shaded curves indicate the two A and B components whose 
area has been used to determine each data point of the photoelectron 
diffraction patterns. Also shown is the spectrum 
taken at a lower coverage (filled circles $\bullet$), close to 
the deposition stage of panel $b$ in Fig.~1), where 
residual $(2 \times 2)$ domains are still present. This spectrum has 
been taken with an overall energy resolution of 120~meV.
  } 
\label{figSn4d}
\end{figure}

Fig.~\ref{figSn4d} shows a typical Sn $4d$ photoemission spectrum taken 
at $\sim 210$~eV photon energy on the $(3 \times 3)$ phase. In 
agreement with previous observations,\cite{lelay,uhrberg,avila99,avila}
but in contrast with the results presented by Kidd et al.,\cite{kidd}
it can be fitted to two 
spin-orbit split doublets, shifted by $0.38 \pm 0.02$~eV and with 
an intensity ratio close to 2:1.  
Any deviation from this shape was always associated to residual 
$(2 \times 
2)$ or $(7 \times 7)$ phases as observed by RHEED patterns. 
A spectrum taken at a deposition 
stage close to that shown in the panel $b$ of 
Fig.~\ref{figRHEED} can also be seen in 
Fig.~\ref{figSn4d}. This spectrum shows a $4d_{5/2}$ peak at 24.3~eV 
binding energy, which is characteristic of the $(2 \times 2)$ phase. This 
observation is fully consistent with the previous 
measurements reported in the literature.\cite{lelay,avila99,avila} 
The experiments have been performed by taking the whole Sn 4d 
spectrum for each point of the PED angular and photon energy scans at 
T$_{s} \sim 150$~K. 
Four gaussians have been used to fit the spectra, 
while both the spin-orbit splitting of the doublets and the energy 
splitting between 
the A and B components were kept fixed. The same full width at half 
maximum has been used for all the components. A low order polinomial 
($n \leq 3$) has been used for background subtraction.
The PED data points shown in the present analysis represent the so 
called $\chi$-function,\cite{woodruff} $\chi = \frac{I-I_{0}}{I_{0}}$, 
where $I$ is 
the integrated intensity
of the $4d_{5/2}$ component of the A or B type Sn 
atoms. $I_{0}$ is a smoothly varying background of $I$, modulating 
both the angular and energy PED features.
$I_{0}$ essentially takes into account the energy dependence of the 
atomic photoionization cross-section and the angular dependence of the 
polarization and of the illuminated sample area. 
$I_{0}$ has been evaluated by a low order ($\leq 5$) polynomial fitting of 
$I$.

\begin{figure}
	\centering
	\includegraphics[width=.44\textwidth]{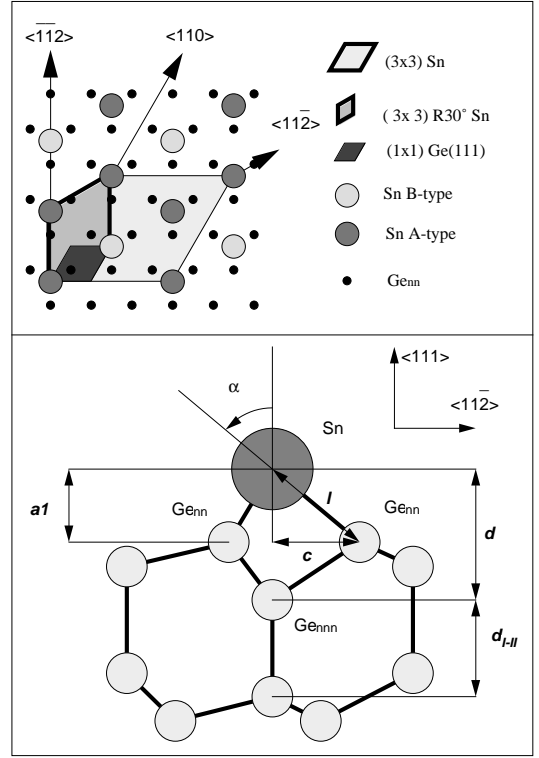}
\caption{ 
Upper panel: a sketch of the $(3 \times 3)$-Sn/Ge(111) 
structure is shown, top view. The underlying matrix of small filled circles 
represents the 
lattice of nearest neighbour Ge$_{nn}$ atoms. The Sn atoms are 
in T$_{4}$ absorption sites. The unit cells 
corresponding to the $(1 \times 1)$-Ge(111), $(\sqrt 3 \times \sqrt 
3)$R30$^{\circ}$-Sn/Ge(111) 
and $(3 \times 3)$-Sn/Ge(111) lattices are indicated by the shaded areas. 
The main symmetry 
directions are also indicated.
Lower panel: a sideview of the Sn atom and its 
closest Ge  atoms. 
The main fitting parameters $\alpha$, $l$, $d$ and $d_{I-II}$ are 
also indicated.
} 
\label{figdrawing}
\end{figure}

\begin{figure}
	\centering
	\includegraphics{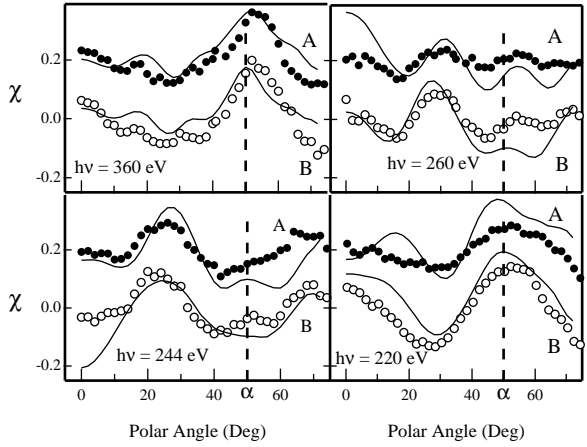}
	\caption{Photoelectron diffraction polar scans taken along the 
$[\overline{1} \overline{1} 2]$ 
 direction at four different photon energies: 220, 244, 260 and 
360~eV. The emission angle is measured from the surface normal 
(0$^{\circ}$). An offset of 0.2 has been added to the 
A $\chi$-function for the sake of clarity. 
At 50-52$^{\circ}$, the scans display a 
maximum (220 and 360~eV) or a minimum (244 and 260~eV) depending on the 
interference conditions of the primary photoelectron wave with that
backscattered by the nearest neighbour Ge atoms. 
The best fit simulations are also shown (full 
lines). The vertical dashed lines indicate the Sn-Ge$_{nn}$ bond 
angle $\alpha$ from the surface normal.
}
	\protect\label{figpolar}
\end{figure}

\begin{figure}
	\centering
	\includegraphics{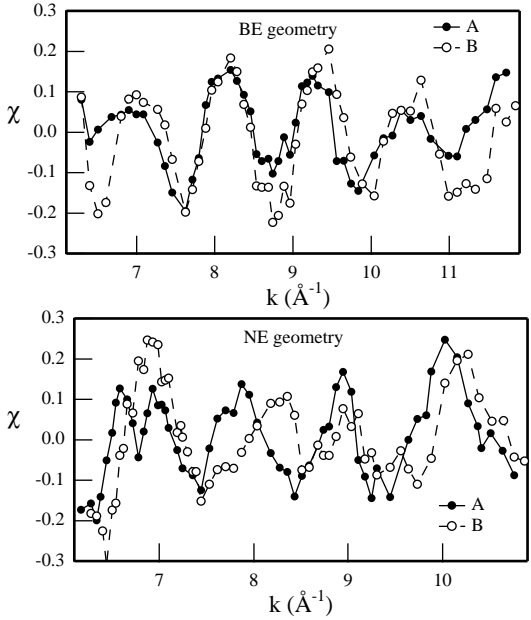}
	\caption{Upper panel: $\chi$-function of the energy scan in bond 
	emission geometry for the A and B components (filled and open 
	circles, respectively). This PED measurement is mainly sensitive to 
	the Sn-Ge$_{nn}$ bond length $l$.
	Lower panel: $\chi$-function of the energy scan in normal 
	emission geometry for the A and B components (filled and open 
	circles, respectively). This PED measurement is mainly sensitive to 
	the Sn vertical height $d$ above the Ge$_{nnn}$ atom underneath.
	Note that the data presented here are based on a more extended set 
	of spectra with respect to our previous study,\cite{floreano} thus 
	yielding a better statistics and, in the case of normal emission 
	data, a more extended energy range too.
	 }
	\protect\label{figchifunction}
\end{figure}

First we have measured the polar scans from 
the surface normal (normal emission, NE, conditions) 
to the horizon (grazing emission) 
for six different photon energies and the three inequivalent main 
symmetry directions of the surface. 
The scans have been performed by keeping the sample in 
transverse-magnetic ($TM$) 
polarization, i.e. with the scattering plane normal to the magnetic 
field of the photon beam, and rotating the analyzer in the 
scattering plane. When 
the surface is oriented with the [${\overline 1}$${\overline 1}$2] 
direction in the scattering plane, 
at a specific angle $\alpha$ from the surface normal the analyzer is aligned 
with the 
Sn-Ge$_{nn}$ bond angle  (see Fig.~\ref{figdrawing}).
In this, 
so called, bond emission (BE) geometry, 
the Sn photoelectrons can be efficiently backscattered by 
the Ge$_{nn}$ atom and give rise to intensity maxima and minima, 
when changing the photon energy. By a simple visual inspection, the bond 
direction $\alpha$ for both A and B Sn atoms can be roughly estimated 
as
50$^{\circ}$ from the surface normal (see the polar scans taken at 360 
and 220~eV in Fig.~\ref{figpolar}). 
Given the information on the bond direction, 
the bond length $l$ has been determined by variable energy PED scans in BE 
geometry. The polarization vector of the photon beam has been oriented at 
50$^{\circ}$ from the surface normal, along the bond direction. 
This choice enhances the 
sensitivity to the Sn-Ge$_{nn}$ bond length, as shown by MSCD 
simulations.\cite{floreano}
A very similar photon energy dependence has been found for both A and B 
components 
in BE condition (see upper panel of Fig.~\ref{figchifunction}).\cite{floreano}
Finally, the Sn height $d$ above the underneath Ge$_{nnn}$ atom 
has been measured by taking energy scans in NE conditions
 and with the sample surface in $TM$ polarization. 
In this case the A and B components display a remarkably 
different energy dependence, as can be seen in the lower panel of 
Fig.~\ref{figchifunction}. 

The quantitative evaluation of $l$, $d$ and $\alpha$ for both A and B 
type of Sn atoms has been 
performed by fitting the data to the MSCD simulations, starting from 
the geometric structure recently obtained by LDA-GC 
calculations.\cite{degironcoli}
 The simulations have been performed on a 30 atom cluster
centered on the Sn atom by considering multiple scattering up 
to the 8th order, 4th Rehr-Albers order and 
with a pathcut of 1$\times 10^{-4}$.\cite{mscd}
Lower pathcut values did not modify the simulations appreciably.
A Debye temperature of 230~K and 
an inner potential $V_{0} = 10$~eV have been obtained by fitting the energy 
scans and successively they have been 
kept fixed in the recursive fitting procedure. 

The fit quality has 
been evaluated by means of the commonly used reliability factor 
$R=\frac{\sum{(\chi _{exp} - \chi _{calc}) ^{2}}}
{\sum{(\chi _{exp} ^{2} + \chi _{calc} ^{2})}}$. 
In order to compare our results with previously published 
structural studies, the random error has been estimated following the 
approach of Woodruff and Bradshaw for the analysis of the 
variable energy PED.\cite{booth}
By applying the method described by Pendry for LEED,\cite{pendry} the 
variance of the R-factor is calculated for its minimum value.
The variance is 
evaluated as $Var(R_{min})= 
R_{min} \sqrt{2/N}$, where $N$ is the number of independent pieces of 
structural information within the spectrum energy range.\cite{booth}
All the parameter values $p_{i}$ yielding structures with associated R-factor 
lower than $R_{min}+Var(R_{min})$ are regarded as falling within one 
standard deviation from that of the best fit structure. 
Assuming a parabolic form of the R-factor,
the random error on the $p_{i}$ parameter is evaluated as 

\begin{equation}
\Delta p_{i} = 
\left( \frac{2 \cdot Var(R_{min})}{\left(\frac{\delta^{2} R}{\delta 
p_{i}^{2}}\right)_{min}}\right)^{1/2},
\end{equation}

where the other $p_{j}$ parameters ($j \neq i$) are taken at the 
minimum of the R-factor.
As a result an error of $\pm 0.03$ and $\pm 0.05$~\AA~ has been 
found for $l$ and $d$, respectively. 
 Due to the intrinsic limit of this approach, that assumes the 
structural parameters to be independent, we have also checked, by 
visual inspection,
that a 
deviation of 0.05~\AA~ from the best fit value
produces significative differences in the simulation curve.

Due to computational limits, the BE and NE energy scans 
have been independently fitted. However, they are 
essentially sensitive to a single parameter ($l$ and $d$, respectively).
The preliminary results for $l$ and $d$, obtained by comparing
data and simulations by visual inspection, have then been used 
to fit all the polar 
scans at the same time in order to obtain the bond angle $\alpha$. 
The new value for $\alpha$ has then been used to obtain better $l$ 
and $d$ values from the R-factor analysis of the energy scans. 
This fitting 
procedure has been iterated until a  self-consistent 
evaluation of $l$, $d$ and $\alpha$ was obtained for both A and B 
components from the entire set of PED data.

\begin{figure}
	\centering
	\includegraphics{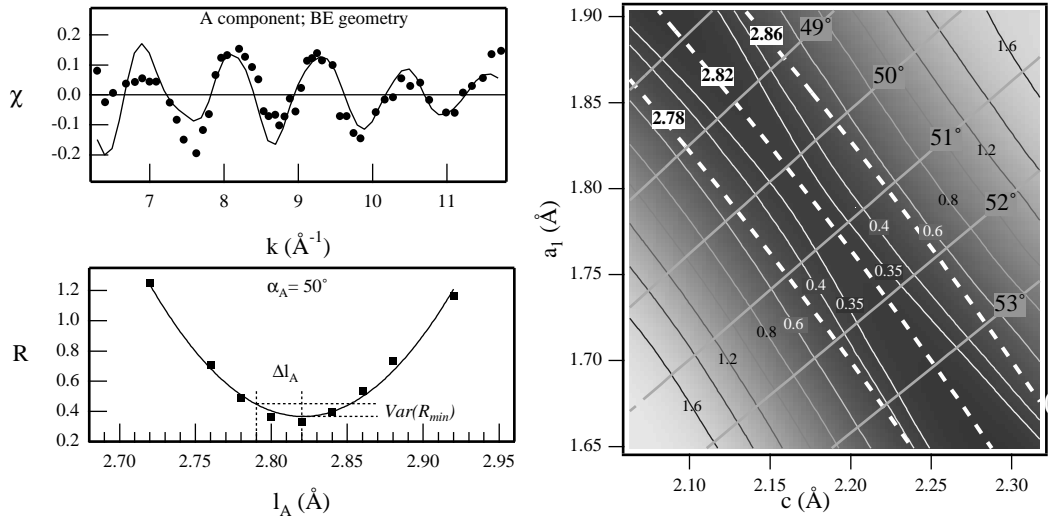}
	\caption{ Bond length analysis of the main component in bond emission (BE)
	geometry. 	 Both the 
	photon polarization vector and the electron analyzer have been 
	oriented 
	at 50$^{\circ}$ from the surface normal, along the Sn-Ge$_{nn}$~ bond
	direction.  
	Left upper panel:  kinetic momentum (wavenumber) dependence of the 
	$\chi$-function for the A (majority) component of the Sn $4d$ 
	photoelectrons 
	in bond emission. The best fit (full line) is overimposed to the 
	experimental data (filled circles $\bullet$). Right panel: contour plot of 
	the MSCD simulation R-factor for the $a_{1}$ and $c$ parameters (see 
	Fig.~3). The bond length $l_{A}$ (dashed thick 
	lines with markers on white background) and the bond angle $\alpha _{A}$  
	(full thick lines with markers on gray background) are
	 reported for a few values.
	 Left bottom panel: statistical analysis of the random error for the 
	 parameter 
	 $l_{A}$ with the other parameters at their best fit value, as 
	 reported in Table~1. The R-factor values (filled squares) have been 
	 fitted to 
	 a parabolic curve (full line). The R-factor variance and the 
	 corresponding error on the parameter are also reported.
	 }
	\protect\label{figBEmain}
\end{figure}

\begin{figure}
	\centering
	\includegraphics{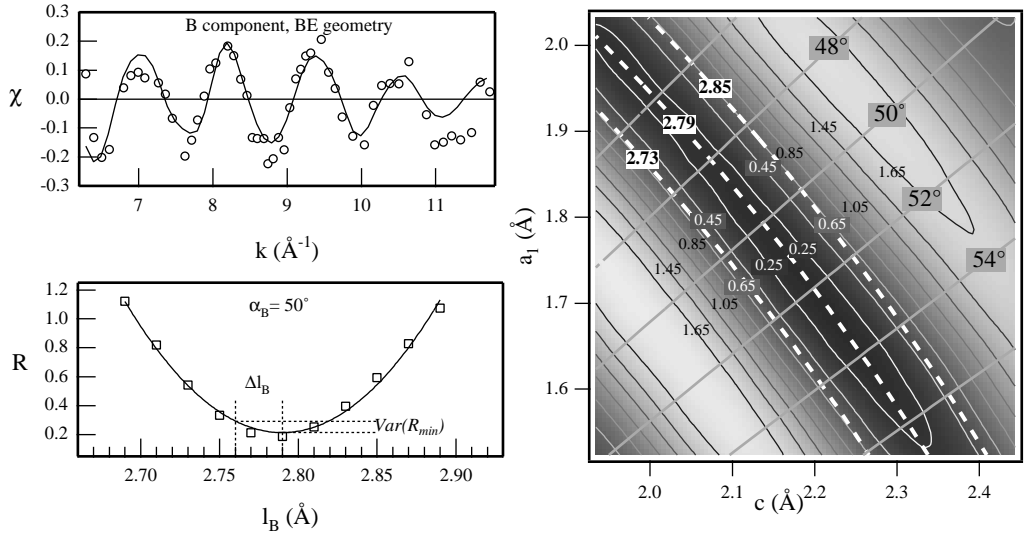}
	\caption{
	 Bond length analysis of the secondary component in BE geometry. The 
	 labeling is the same of Fig.~6.
	 }
	\protect\label{figBEsec}
\end{figure}

The refined bond length $l$ and its error have been found from the 
BE data by calculating the reliability 
R-factor for grid simulations on two parameters, $a_{1}$ and 
$c$ ($l^{2}=a_{1}^{2}+c^{2}$, see Fig.~\ref{figdrawing}) with the 
other parameters fixed at their best fit value. As can be seen 
in Figs.~\ref{figBEmain} and 
\ref{figBEsec},
a well defined minimum is obtained for $l_{A} = 2.82 \pm 0.03$~\AA~ 
and $l_{B} = 2.79 \pm 0.03$~\AA, while an 
uncertainty of almost 10$^{\circ}$ is given for the bond direction.
 We have also observed that, in BE 
conditions, the simulation is weakly affected by the 
position of the
surrounding next nearest neighbour Ge atoms (not even by the Ge$_{nnn}$ 
underneath the Sn atom).

\begin{figure}
	\centering
	\includegraphics{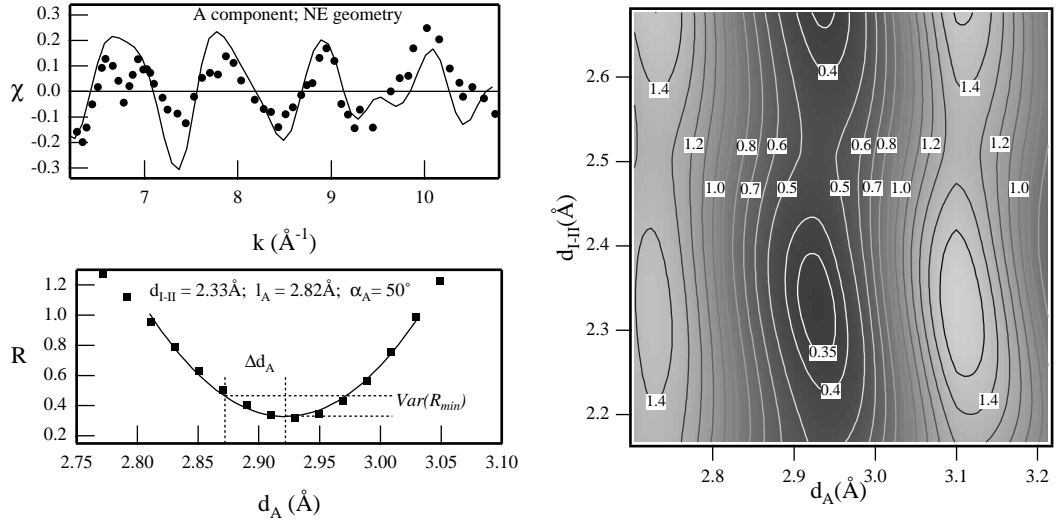}
	\caption{
	Vertical height analysis of the main component  in normal emission (NE)
	geometry.  Measurements taken in transverse-magnetic polarization and with the
	analyzer oriented along the surface normal. 
	Left upper panel:  kinetic momentum (wavenumber) dependence of the 
	$\chi$-function for the A (majority) component of the Sn $4d$ photoelectrons 
	in normal emission. The best fit (full line) is overimposed to the 
	experimental data (filled circles $\bullet$). Right panel: contour plot of 
	the MSCD simulation R-factor for $d_{A}$ and $d_{I-II}$ parameters (see 
	Fig.~3); the R-factor values are reported on a white foreground.
	 Left bottom panel: statistical analysis of the random error for the parameter 
	 $d_{A}$ with the other parameters at their best fit value, as 
	 reported in Table~1. The R-factor values (filled squares) have been fitted to 
	 a parabolic curve (full line). The R-factor variance and the 
	 corresponding error on the parameter are also reported.
	 }
	\protect\label{figNEmain}
\end{figure}

\begin{figure}
	\centering
	\includegraphics{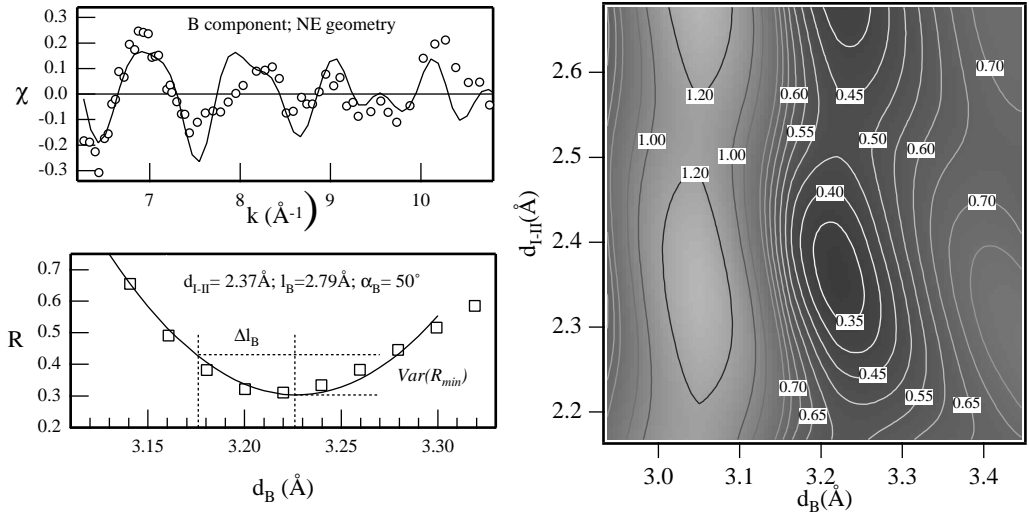}
	\caption{
	 Vertical height analysis of the  secondary component in normal emission (NE) 
	 geometry. The labeling is the same of Fig.~8.
	 }
	\protect\label{figNEsec}
\end{figure}

In the case of the NE scan, the grid scan simulations have been 
performed spanning over the Sn-Ge$_{nnn}$ distance $d$ and the 
distance $d_{I-II}$ between the first and second Ge bi-layers. 
 As shown in Figs.~\ref{figNEmain} and \ref{figNEsec}, 
the R-factor analysis yields two different 
heights, $d_{A} = 2.92 \pm 0.05$~\AA~ and $d_{B} = 3.23 \pm 0.05$~\AA, 
for the A (majority) and B (minority) type Sn atoms. 
It must be emphasized that the cluster used for the 
simulations contains  one Sn atom only. 
The relative 
height between the A and B clusters is thus undetermined, but the 
difference  $d_{B}-d_{A}$ of $\sim 0.3$~\AA~ is in good agreement 
with the calculated\cite{degironcoli,ortega} and 
measured\cite{bunk,zhang}
 vertical ripple of the adatom layer. For what concerns other 
structural parameters, the NE scans have been found weakly sensitive 
to $d_{I-II}$. 
From a comparison with the polar scan simulations, $d_{I-II}$ has 
been estimated to be 0.05-0.1~\AA~ shorter than the Ge bulk interlayer 
separation, for both A and B type Sn atoms.

The polar scans are sensitive to the whole atomic environment of the 
Sn atom (particularly to the bond angle $\alpha$), 
but they display much less features with respect 
to the energy scan (see Fig.~\ref{figpolar}). 
The angular patterns have thus been fitted to 
check the consistency of the overall structure, as obtained by the NE 
and BE scans. The same bond 
angle $\alpha  = 50^{\circ}$
has been determined for both A and B type Sn atoms, with an
 uncertainty of $\pm 3^{\circ}$.

The parameters obtained from the analysis of the whole PED 
data set are shown in Table~\ref{table1} and compared to the 
calculated\cite{degironcoli,ortega} and previously 
measured\cite{bunk,zhang} 
values.
There is a qualitative agreement concerning the
overall picture. In particular, the nearest neighbour 
Ge$_{nn}$ atoms nearly
follow the Sn atom in its vertical displacement. 
According to our data, the three nearest neighbour Ge$_{nn}$ atoms 
 have the same bond length and direction for both A and B type Sn 
 atom, within $\pm 0.03$~\AA~ and $\pm 3^{\circ}$ respectively.
 This is also consistent with the ripple of the Ge atoms of the first 
 layer as reported in the literature ($\sim 
 0.1-0.2$~\AA).\cite{bunk,zhang,degironcoli,ortega}
The most recent DFT calculations\cite{degironcoli,ortega} are in agreement 
with our results, particularly 
for what concerns the vertical height $d$. Small discrepancies are 
only obtained for the Sn-Ge$_{nn}$ bond
angle, which, on the other hand, yields the higher uncertainty 
in our measurements.
Therefore the PED of the two components of the Sn $4d$ spectrum, 
characteristic of the $(3 \times 3)$ phase,\cite{uhrberg,avila} 
yields a  vertically 
distorted structure, consistently with the XRD\cite{bunk} and 
XRD/LEED\cite{zhang} determinations. The component with the highest 
binding energy (minority B) is associated to the Sn atoms that 
protrude above the surface.

\section{Conclusions}

By comparing the RHEED patterns and the photoemission spectra we have 
checked that the correct lineshape of the Sn $4d$ spectrum on the $(3 
\times 3)$ phase is similar to that reported in several 
articles.\cite{lelay,uhrberg,avila99,avila}
In addition, the lineshape, recently proposed by Kidd et al. for the $(3 
\times 3)$ phase,\cite{kidd} has been only observed when residual $(2 \times 2)$ 
phase was present.
We have determined the structure of the low temperature $(3 \times 
3)$-Sn/Ge(111) phase by means of photoelectron diffraction from the Sn 
$4d$ core levels.  The PED measurements and analysis have been performed 
on a set of 
photoemission spectra, which can be always fitted to two components.
Exploiting 
the scattering geometry of the experimental apparatus, we have separately 
measured the Sn-Ge$_{nn}$ bond parameters and the vertical height of 
Sn above the underneath Ge$_{nnn}$ atom for the two kind of Sn adatoms. 
The results are consistent with the 
most recent XRD measurements\cite{bunk} and theoretical 
predictions\cite{degironcoli,ortega}, where one Sn 
atom out of three is vertically displaced by $\sim 0.3$~\AA.
This vertical ripple is strongly coupled to the distortion of
 the first Ge bi-layer.
In particular, the tetrahedron formed by the Sn atom and its 
three nearest neighbours Ge$_{nn}$ atoms is found to retain
 the same structure for both A 
and B components within the experimental error.
The vertical height above the Ge$_{nnn}$ atom is the main difference 
between the two bonding geometries. The Sn atoms that protrude out of 
the surface have the highest binding energy (minority component), 
while the majority A component (lowest 
binding energy) is associated to the Sn atoms that are pushed closer 
to the Ge(111) surface. 

\section{acknowledgements}

We are grateful to M.A. Van Hove and Y. Chen for providing the MSCD 
package for PED analysis. We thank E. Tosatti for useful discussions.
This research has been partially funded by MURST cofin99 (Prot. 
9902332155), by Regione Friuli-Venezia Giulia 98 and by INFM-PAIS F99.
M.B. acknowledges a grant by MURST cofin99 (Prot. 9902112831).

Added note: meanwhile the present paper was being refereed, a 
photoemission and LEED study was published, showing the correlation 
between the Sn $4d$ line shape and the (3$\times$3)-Sn/Ge(111) phase, 
in full agreement with our analysis.\cite{uhrberg2}


\newpage
\narrowtext
\onecolumn

\begin{table}
\centering 
\begin{tabular}{cccccc}
Parameter & SXRD\cite{bunk} & SXRD/LEED\cite{zhang} & LDA-LO\cite{ortega} & LDA-GC\cite{degironcoli} 
& Present paper\\
\hline
$\mathrm{\alpha_{A}}$ & 50$^{\circ}$ & 51.5$^{\circ}$ & 51.8$^{\circ}$ & 51.2$^{\circ}$ & 50$^{\circ}$ \\
$\mathrm{\alpha_{B}}$ & 47.4$^{\circ}$ & 46.9$^{\circ}$ & 49$^{\circ}$ & 47$^{\circ}$ & 50$^{\circ}$ \\
$\mathrm{l_{A} (\AA)}$ & 2.80 & 2.89 & 2.78 & 2.79 & 2.82 \\
$\mathrm{l_{B} (\AA)}$ & 2.88 & 2.93 & 2.79 & 2.84 & 2.79 \\
$\mathrm{d_{A} (\AA)}$ & 3.03 & 2.97 & 2.90 & 2.92 & 2.92 \\
$\mathrm{d_{B} (\AA)}$ & 3.32 & 3.34 & 3.13 & 3.25 & 3.23 \\
$\mathrm{d_{I-II} (\AA)}$ & 2.36-2.39 & 2.39 & 2.31-2.36 &  2.34-2.39 &  2.34-2.40 \\
$\mathrm{\Delta Sn (\AA)}$& 0.29 & 0.37 & 0.23 &  0.33 & 0.31 \\
\end{tabular}
\caption{\small The best fit values for the parameters $\alpha$, 
$l$, $d$ and $d_{I-II}$, compared to recent experimental measurements 
and theoretical calculations. The calculated random errors are $\Delta 
l = 0.03$~\AA, $\Delta 
d = 0.05$~\AA, $\Delta 
d_{I-II} = 0.12$~\AA, and $\Delta 
\alpha = 3^{\circ}$ for both A and B components. 
The vertical ripple $\Delta$Sn has been simply 
evaluated as $d_{B}-d_{A}$. The geometric parameters reported for the 
majority A component from Ref.~[5] represent the mean value of the two 
lower height Sn atoms (whose coordinates were separately 
determined). 
}
\protect\label{table1}
\end{table}

\twocolumn

\onecolumn

\end{document}